# Size Dependence in Multicolor Upconversion in Single $Yb^{3+}$, $Er^{3+}$ Co-doped $NaYF_4$ Nanocrystals

*Peter A. Tanner[†,*] and Chang-Kui Duan[¶]*

Department of Biology and Chemistry, City University of Hong Kong, Tat Chee Avenue, Kowloon, Hong Kong S.A.R., Peoples' Republic of China, and Institute of Modern Physics, Chongqing University of Post and Telecommunications, Chongqing 400065, Peoples' Republic of China

.

AUTHOR EMAIL ADDRESS bhtan@cityu.edu.hk

ABSTRACT

Schietinger et al. (*Nano Lett.* **2009**, *9*, 2477-2481) observed single nanoparticle emission from NaYF$_4$:Er$^{3+}$,Yb$^{3+}$ and noted unusual changes in the intensities of red and green emissions with particle size. We comment that the documented changes may be due to instrumental artifacts and that the explanations given for the changes are incorrect.

KEYWORDS Upconversion, Size-dependence, Erbium, Ytterbium, NaYF$_4$, Nanocrystal

Schietinger et al.[1] recently utilized an atomic force microscope coupled with a spectrometer to study the upconversion in the visible spectral region of α-NaYF$_4$ crystals co-doped with Yb$^{3+}$ and Er$^{3+}$. This system is one of the most efficient upconverters of infrared to visible radiation,[2] and it especially gives green (500-580 nm) and red (620-720 nm) radiation when excited by a diode laser at 960-980 nm. Although previous studies have largely focused upon the more efficient hexagonal phase (β-NaYF$_4$) rather than the cubic one (α-NaYF$_4$), most attention for the latter has been devoted to monodisperse nanparticles,[3,4] white light generation[5] and multicolor fine-tuning[6] of nanoparticles, and their size-control.[7] The study of Schietinger et al.[1] reported an unusual experimental result for α-NaYF$_4$ nanoparticles and gave an explanation for the result. It is the purpose of this communication to demonstrate that the experimental result is an artifact and the explanation is incorrect.

In the experiment of Schietinger et al.,[1] the intensity ratio of the green to red emission (green-to-red ratio: GRR) was found to increase by a factor of nearly 4 when the particle size of the nanocrystals was reduced from 65 nm to 5.6 nm. It is well known that many parameters can change the GRR, and these include the incident laser power, temperature,[8] dopant ion levels,[9] particle shape,[10] phase purity and the presence of other impurities. It is accepted that these parameters were held fairly constant in the experiments of Schietinger et al. A previous study of another Yb$^{3+}$,Er$^{3+}$ co-doped system, Y$_2$O$_3$,[11] has found that decreasing the particle size leads to a smaller GRR ratio. This is readily explained by relatively greater multiphonon relaxation rates from the $^4S_{3/2}$ and $^4I_{11/2}$ multiplet terms in smaller nanoparticles. These particles have a reduced surface to volume ratio so that a larger proportion of Er$^{3+}$ ions are nearer to the surface and under the influence of high energy vibrations from contaminants such as $H_2O$, $OH^-$, etc. The findings of Schietinger et al.[1] are therefore unexpected. In other reported studies of α-NaYF$_4$ nanoparticles with the similar composition (Yb 20-25 at.%; Er 2 at.%), the GRR ratio was measured as 0.45 for 20 nm particles[6]; 0.31 for 14 nm particles[4]; and 0.60 for 20 nm particles,[3] although in each case there is a dispersion of particle size. The GRR ratio for a single 47 nm particle measured by Schietinger et al. was between 0.54 to 0.76 (Figs. 2,3 in Ref. 1), which is consistent with the above ratios considering that other factors mentioned above could differ.

The order of a multiphoton process with moderate laser power[12] can be determined from a log-log plot of upconversion intensity and incident laser power. It has been demonstrated in many previous studies of $Yb^{3+}$, $Er^{3+}$ upconversion that both of the emissions from the $Er^{3+}$ multiplets $^4S_{3/2}$ and $^4F_{9/2}$ are two-photon processes,[4] but as clarified by Pollnau et al.,[12] the slope of the log-log plots can vary between 2 (in the limit of infinitely small upconversion) to 1 (in the limit of infinitely large upconversion rates). Other complications which can lead to a change in slope from the value of 2 are: (i) overlapping of the $^4S_{3/2} \rightarrow {}^4I_{15/2}$ (Green, with maximum intensity around 555 nm) transition with $^2G(H)_{9/2} \rightarrow {}^4I_{13/2}$ (peaking around 555 nm): this may readily be corrected since the oscillator strength of the latter transition is ~2.1 times that of $^2G(H)_{9/2} \rightarrow {}^4I_{15/2}$ (peaking around 410 nm) [13]; (ii) overlapping of the emission $^4F_{9/2} \rightarrow {}^4I_{15/2}$ (Red, peaking around 670 nm ) with $^2G(H)_{9/2} \rightarrow {}^4I_{11/2}$ (peaking around 700 nm): again, this can be corrected for, since the oscillator strength of the latter transition is ~0.78 times that of $^2G(H)_{9/2} \rightarrow {}^4I_{15/2}$ [13]; as well as other factors which can occur at high laser powers but which are not considered here.[14] In the experimental regime of Schietinger et al., moderate laser powers of 25 mW and 100 mW were employed. Under these conditions, the slopes of log-log plots of intensity vs laser power can be derived from Figs. 2 and 3 in their work for the 47.1 nm particle as 1.8 for Green and 0.5 for Red; and for the 5.6 nm particle as 0.75 for Green and 0.43 for Red. Clearly there is an error in intensity measurements for these particles and it is particularly discriminating against the red emission for the 5.6 nm particle. We therefore do not consider that the GRR values given in Fig. 4 of Schietinger et al. are accurate, and do not provide a valid variation with particle size.

In order to justify the anomalous surge of intensity of Green emission, compared with Red, with decreasing particle size, Schietinger et al. suggested that the change in GRR was due to a size-related bottleneck for nonradiative phonon relaxation processes. The authors considered that for the smaller particles, not enough phonons are present to efficiently populate the red-emitting level. We consider these explanations to be erroneous. The energy gaps below ($^2H_{11/2}$,$^4S_{3/2}$) and $^4F_{9/2}$ were given as 2000 cm$^{-1}$ and 5000 cm$^{-1}$.

It is well-known that the phonon density of states is modified in nanoparticles, with respect to bulk systems. For example, this modification has a profound effect upon the low-frequency modes[15,16] and changes the band shapes of vibrations in the Raman spectra.[17] The multi-phonon relaxations in lanthanide ions concerned here involve the higher energy phonon modes which are often considered approximately as localized modes and these are less-affected by particle-size changes than the lowest energy modes. The fundamental misconception in the explanation of Schietinger et al. is that phonons are required for multiphonon relaxation in these cases. In fact, phonons are created due to the conservation of energy and this does not depend on the availability of phonons but on the availability of modes and coupling between electrons and vibrations.

In the case of 4f$^N$-4f$^N$ transitions, the electron-phonon coupling is weak and the multiphonon relaxation rate $W_{NR}(\Delta E,T)$ for low temperature $T \sim 0$ is due to spontaneous emission of phonons and can be described by the energy gap law:[18]

$$W_{NR}(\Delta E,0) \sim W_0 \exp(-\alpha\, \Delta E /\hbar\omega_{max}), \qquad (1)$$

where $\alpha$ and $W_0$ may depend upon the electron-phonon coupling constant. At higher temperature $T$, induced phonon emission may be important. The $N$-phonon relaxation rate depends on the number of phonons per mode $<n(T)> = 1/(\exp(\hbar\omega_{max}/kT)-1)$ as $(1+<n(T)>)^N$, where $N = \Delta E/\hbar\omega_{max}$.[19]

Since the 4f orbitals of lanthanide ions are very localized, the electron-phonon coupling is expected to be constant for the two nanoparticles, and the effective maximum optical phonons energy $\hbar\omega_{max}$ are also hardly affected by particle size, we do not expect substantial changes in multiphonon relaxation rates.

There is another point. Due to the complex dynamics involved in the populations of $^4F_{9/2}$ and ($^2H_{11/2}$, $^4S_{3/2}$), we cannot simply consider the energy gap involved in the upconversion processes as 2000 cm$^{-1}$ and 5000 cm$^{-1}$. As shown by Suvyer et al.,[20] at room temperature, $^2H_{11/2}$ and $^4S_{3/2}$ are populated due to the relaxation of $^4F_{7/2}$ to $^2H_{11/2}$ (and subsequently to $^4S_{3/2}$ for the latter). The energy gap for multiphonon relaxation is only ~1200 cm$^{-1}$ and less than the energy of 5 phonons and so the relaxation is expected to be very efficient. The multiplet $^4F_{9/2}$ is populated by either the relaxation of $^4S_{3/2}$,[4,20] with an energy gap of 2900 cm$^{-1}$ or subsequent energy transfer excitation of $^4I_{13/2}$,[20] which is populated by relaxation of $^4I_{11/2}$, with an energy gap of 3500 cm$^{-1}$. These two gaps are much more than 5 phonons and the multiphonon relaxations are generally inefficient but levels may be more efficiently populated by cross relaxations,[14] which are generally near resonant for the concentration considered and room temperature.[20] In those cross relaxations, concentrations and thermal population of some levels are important, while the gaps, maximum phonon energy and number of phonons etc. are less relevant.

**Acknowledgements**

Financial support for this work from the Hong Kong Research grants Council General Research Fund Grant CityU 102308 is gratefully acknowledged. This work was also supported by the National Science Foundation of China, under Grant No. 10874253.

This paper was submitted to Nano Lett. but was returned by the Editor without review. A subsequent appeal was unanswered.